# Toward Youth-Centered Privacy-by-Design in Smart Devices: A Systematic Review

Molly Campbell
*Computer Science Department*
*Vancouver Island University*
Nanaimo, Canada
molly.campbell@viu.ca

Mohamad Sheikho Al Jasem
*Computer Science Department*
*Vancouver Island University*
Nanaimo, Canada
mohamad.sheikhoaljasem@viu.ca

Ajay Kumar Shrestha
*Computer Science Department*
*Vancouver Island University*
Nanaimo, Canada
ajay.shrestha@viu.ca

***Abstract*— This literature review evaluates privacy-by-design frameworks, tools, and policies intended to protect youth in AI-enabled smart devices using a PRISMA-guided workflow. Sources from major academic and grey-literature repositories from the past decade were screened. The search identified 2,216 records; after deduplication and screening, 645 articles underwent eligibility assessment, and 122 were included for analysis. The corpus was organized along three thematic categories: technical solutions, policy/regulatory measures, and education/awareness strategies. Findings reveal that while technical interventions such as on-device processing, federated learning, and lightweight encryption significantly reduce data exposure, their adoption remains limited. Policy frameworks, including the EU's GDPR, the UK Age-Appropriate Design Code, and Canada's PIPEDA, provide important baselines but are hindered by gaps in enforcement and age-appropriate design obligations, while educational initiatives are rarely integrated systematically into curricula. Overall, the corpus skews toward technical solutions (67%) relative to policy (21%) and education (12%), indicating an implementation gap outside the technical domain. To address these challenges, we recommend a multi-stakeholder model in which policymakers, manufacturers, and educators co-develop inclusive, transparent, and context-sensitive privacy ecosystems. This work advances discourse on youth data protection by offering empirically grounded insights and actionable recommendations for the design of ethical, privacy-preserving AI systems tailored to young users.**

## I. Introduction

The rapid adoption of AI-driven smart devices among youth (16-24) has revolutionized connectivity but introduced privacy risks, from unauthorized data access to behavioral surveillance [1], [2]. While IoT technologies like wearables and smart home systems offer convenience, their design often prioritizes functionality over security, leaving young users vulnerable. Studies reveal persistent gaps in youth privacy protections, and technical measures like encryption and federated learning (FL) face scalability challenges [3], [4]. Policy frameworks such as Canada's Personal Information Protection and Electronic Documents Act (PIPEDA) lack youth-specific enforcement mechanisms [5]. Ethical concerns persist, as many smart devices collect sensitive data through opaque data collection processes, as seen in the "privacy paradox", where youth prioritize convenience over privacy despite high-risk awareness [6], [7].

Privacy-by-design solutions aim to bridge these gaps by embedding protection into technological development [8]. Technical innovations, such as decentralized architecture and adaptive controls, show promise but require infrastructure investments and standardization [9]. Policy efforts, like the SANIJO framework for ethical IoT behavior, highlight the need for robust data protection policies [10]. Education remains a critical pillar, though current strategies often overlook the cultural and relational dimension of youth privacy [1], [11]. Resilience-based interventions, such as privacy literacy programs, could empower youth, but their scalability depends on collaboration among policymakers, manufacturers and educators [12], [13], [14], [15].

This systematic review evaluates privacy-by-design solutions aimed at safeguarding youth privacy in smart devices, analyzing peer-reviewed and grey literature across technical, policy, and educational domains. We scope risks across home, school, and health contexts, where assets include ambient audio/video, location, and biometrics, and adversaries include vendors, data brokers, peers, and external attackers. We identify key challenges, including implementation gaps, and the trade-offs between usability and security, while synthesizing interdisciplinary research to inform a safer digital ecosystem. Our findings aim to equip policymakers, technologists, and educators with actionable recommendations, bridging the gap between privacy principles and real-world implementation. Ultimately, this work contributes to building a digital environment where youth can benefit from smart technologies without compromising their privacy.

The remainder of the paper is structured as follows: Section II provides the background and key definitions. Section III outlines the methodology, including the inclusion criteria and data extraction process. Results are presented in Section IV, followed by a discussion of implications and limitations in Section V. Section VI concludes with recommendations for future research and policy.

## II. Background

This section frames the systematic review and informs how to design AI-enabled smart devices with robust, youth-tailored privacy protection. Our contribution is a youth-centered scoping of privacy-by-design patterns across technical, policy, and education strands, complementing broader IoT security surveys.

## A. AI-enabled Smart Devices (IoT)

AI-enabled smart devices integrate intelligence to automate tasks, personalize experiences, and enhance connectivity across wearables, smart assistants, toys, and educational tools [16]. They use machine learning and real-time data processing to adapt to behavior, offering benefits for youth (e.g., interactive learning, health monitoring) alongside risks from weak encryption and unauthorized data sharing [17], [18], [19]. As adoption grows, stakeholders must balance innovation with safeguards such as federated learning and policy-aligned design [20], [21].

## B. Data Privacy and Security Risks for Youth

Youth aged 16-24, spanning late adolescence through the typical post-secondary transition, face substantial risks as smart devices collect extensive personal data with limited transparency or control [1], [2]. Continuous capture by smart speakers, wearables, and connected-home systems exposes routines, behaviors, and preferences [22]. Functionality often outpaces security, enabling surveillance and breaches, compounded by weak encryption and insecure protocols [16], [23], [24]. These conditions facilitate profiling, targeted advertising, and opaque algorithmic decisions [25], while AI can infer sensitive details from seemingly benign signals [26], [27]. Policy gaps further leave youth vulnerable, as existing regulations rarely address their specific needs [18], [28], [29]. The "privacy paradox" intensifies exposure as youth trade privacy for convenience, especially in classrooms using monitoring/AI tools with questionable consent and on social platforms that reward oversharing [6], [18], [30]. Without stronger safeguards, long-term digital autonomy of youth is at risk.

## C. Current Regulatory Landscape

Canada's PIPEDA remains principle-based and neutral, lacking youth-specific safeguards and relying on "meaningful consent," which overestimates young users' capacity to parse complex practices [28], [31]. Unlike stricter frameworks, it does not mandate privacy-by-design technical measures, leaning on industry self-regulation [32]. In contrast, the EU's General Data Protection Regulation (GDPR) and the UK's Age-Appropriate Design Code embed defaults such as data minimization and geolocation limits; GDPR requires parental consent under 16, while the UK code covers users under 18 [21], [33]. PIPEDA's review committee has recommended reforms like a "right to erasure" for minors, but has not codified them into law [28]. Empirically, 59% of IoT apps violate child-specific rules via third-party sharing, underscoring consent limits [16]. Evidence favors hybrid approaches: policy mandates, technical standards, and user education, to close enforcement gaps, especially for cross-border data flows [16], [28], [34]. Without reforms, PIPEDA will be unable to keep up with evolving privacy threats. Policymakers must balance innovation with robust safeguards to ensure privacy-by-design keeps pace with smart-device risks.

# III. METHODOLOGY

We employed Preferred Reporting Items for Systematic reviews and Meta-Analyses (PRISMA)-guided systematic review as shown in Fig. 1 to synthesize research on privacy-by-design frameworks, tools, and policies for youth in smart devices [35]. Child-specific legal obligations (e.g., parental consent) are treated as boundary conditions and are not generalized to legal adults; where relevant, we distinguish minors from adults within the 16-24 cohort. We adopted a scoping emphasis; findings are synthesized qualitatively and should not be interpreted as effect-size estimates. This review evaluated effectiveness, implementation challenges, and actionable recommendations via structured data collection, screening, and thematic analysis.

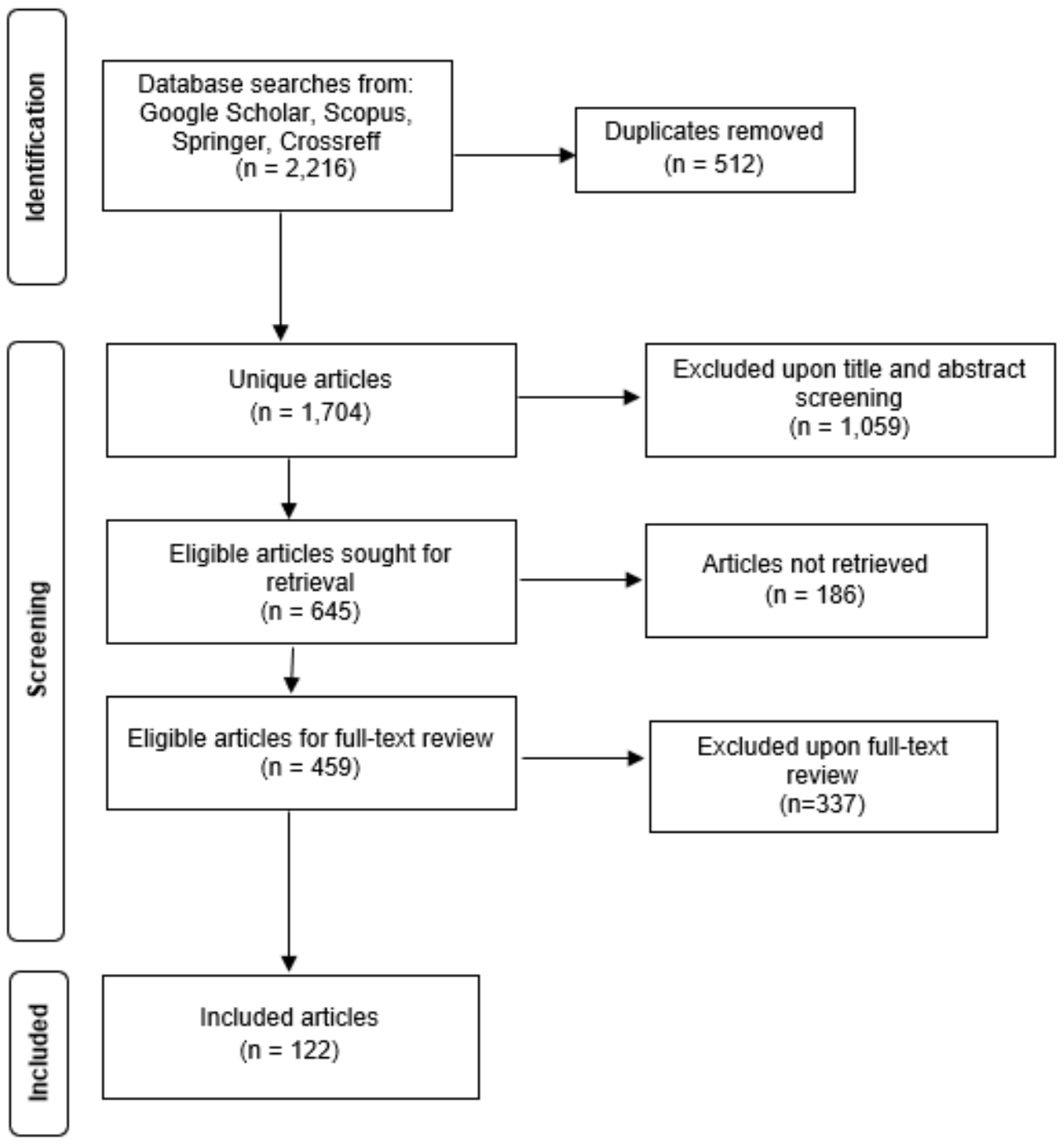

Fig. 1. PRISMA Flow Diagram

## A. Research Question

Central question: *What existing and emerging privacy-by-design solution effectively protects youth in smart devices?* To investigate this, the following questions were formulated:

- *RQ1:* What technical approaches are most effective in enhancing youth data security in smart devices?
- RQ2: How do policy frameworks (e.g., PIPEDA) address youth privacy risks, and what gaps exist in enforcement and compliance?
- RQ3: What role do educational strategies play in improving youth awareness and control over their data?
- RQ4: What are the key challenges in implementing privacy-by-design measures at scale for youth-focused smart devices?
- RQ5: How can policymakers, manufacturers, and educators collaborate to strengthen youth privacy protections?

## B. Literature Collection

We conducted a systematic search across four databases (Google Scholar, Scopus, Springer, Crossref) to identify studies on privacy-by-design solutions for youth in smart devices. We used keywords such as "privacy-by-design in smart devices", "youth privacy policies", "age-appropriate privacy measures", "youth protection in IoT", "Privacy-preserving smart devices", "digital literacy and youth privacy awareness", "Canadian privacy laws and youth data governance", and "smart device processing for privacy", combined with Boolean operators (AND, OR, NOT). We limited the search to the past decade, peer-reviewed articles, conference proceedings, and authoritative grey literature such as government reports, in English only. Grey literature was included to surface emerging practice; pre-print sources were referenced for context rather than for core quantitative claims. For coverage, we implemented a dual approach: automated collection using custom Python scripts that query database APIs for keyword combinations, and supplemented manual searching to ensure relevant material was not missed. The scripts used for this process and the list of reviewed papers are available in the GitHub repository's main branch at commit 1504dc2d9fc8a801e25787d86e2581c5be66009c [36].

## C. Screening and Selection

The initial identification phase involved comprehensive database searches that identified 2,216 potential studies. After removing duplicates using Excel's "Remove Duplicates" function, 1,704 papers remained. Title and abstract screening against the research questions excluded 1,059 records, leaving 645 for eligibility assessment. We included studies that examined privacy-by-design in smart devices, addressing privacy concerns through technical solutions, policy frameworks, or education interventions. Exclusions comprised work not focused on smart devices/IoT, lacking youth relevance, not peer-reviewed, non-English, or with full text unavailable. The final phase involved comprehensive full-text reviews of all papers marked as "Include" during the eligibility assessment. Full-text appraisal verified compliance and designated qualifying articles as "Final Include," resulting in 122 studies.

## D. Extraction of Findings

For each of the included studies, we systematically extracted key findings to address our research questions, documenting study objectives, methodologies, characteristics, and significant outcomes. This extracted data was organized in a structured Excel spreadsheet to enable cross-study analysis and thematic identification.

## E. Synthesis of Findings

The extracted findings were systematically analyzed to identify predominant themes, emerging patterns, and notable contradictions across the literature. Papers were labeled based on their relevance to three primary categories: technical solutions (TS), policy frameworks (P), and education and awareness strategies (EA). For proportional reporting, each study received a single primary thematic label; secondary labels were retained for cross-referencing but are not reflected in Fig. 2. Findings were organized into Excel worksheets by research question, and studies appear on multiple sheets when relevant. This process revealed consistent evidence as well as critical gaps. Given heterogeneity across study designs, a formal risk-of-bias appraisal was out of scope; conclusions are interpretive and qualitatively triangulated.

## F. Paper Writing

We consolidated findings by research question, identified gaps, and derived targeted recommendations. The review was then drafted to interrogate the evidence and analyze those gaps. This approach provides scholarly rigor and practical relevance, advancing understanding of effective privacy protection for youth in increasingly connected device ecosystems.

# IV. Results

This review paper synthesizes privacy-by-design solutions for youth in smart devices across three dimensions: technical mechanisms (e.g., encryption, edge computing), policy and regulatory measures, and education for privacy awareness. Although not all sources exclusively target youth, we retained studies addressing youth-adopted devices (e.g., smartphones, wearables, smart home systems). Thematic analysis addresses the research questions from Section III and surfaces effective approaches alongside persistent gaps, underscoring the need for more youth-specific evidence. Across the included corpus, Technical Solutions account for 67% of studies, Policy Measures 21%, and Education and Awareness Strategies 12% (Fig. 2). This distribution contextualizes the depth of evidence discussed in this section and motivates the emphasis on scaling policy and education efforts.

## A. RQ1: What technical approaches enhance youth data security?

The literature reports a broad set of privacy-preserving techniques spanning encryption, distributed/edge processing, and secure architectures. FL enables collaborative model training without centralized data exposure and is effective for intrusion detection and DDoS mitigation [20], [37], [38]; Proof of Authentication (PoAh)-enabled FL attains 99% accuracy while preserving privacy [39]. Heterogeneous FL (e.g., pseudo-metapath, adaptive multi-dimensional FL) improves performance under non-IID data [40], [41]. Blockchain-enhanced FL ensures data integrity [42], [43], while model-fusion approaches (PrivFusion, FedSWP) combine differential privacy and transfer learning for secure IoT monitoring [44], [45]. Certificateless authentication [46] and privacy-preserving truth discovery [47] reinforce distributed learning, relevant to youth-focused educational and home devices [48], [49], [50], [51]. Functional encryption (e.g., NPANN) [52] demonstrates masked matrices and inner-product encryption for private neural network training, and hybrid binary neural networks (BNNs) [53] and homomorphic encryption (HE) schemes mitigate edge vulnerabilities and membership-inference risks with efficiency.

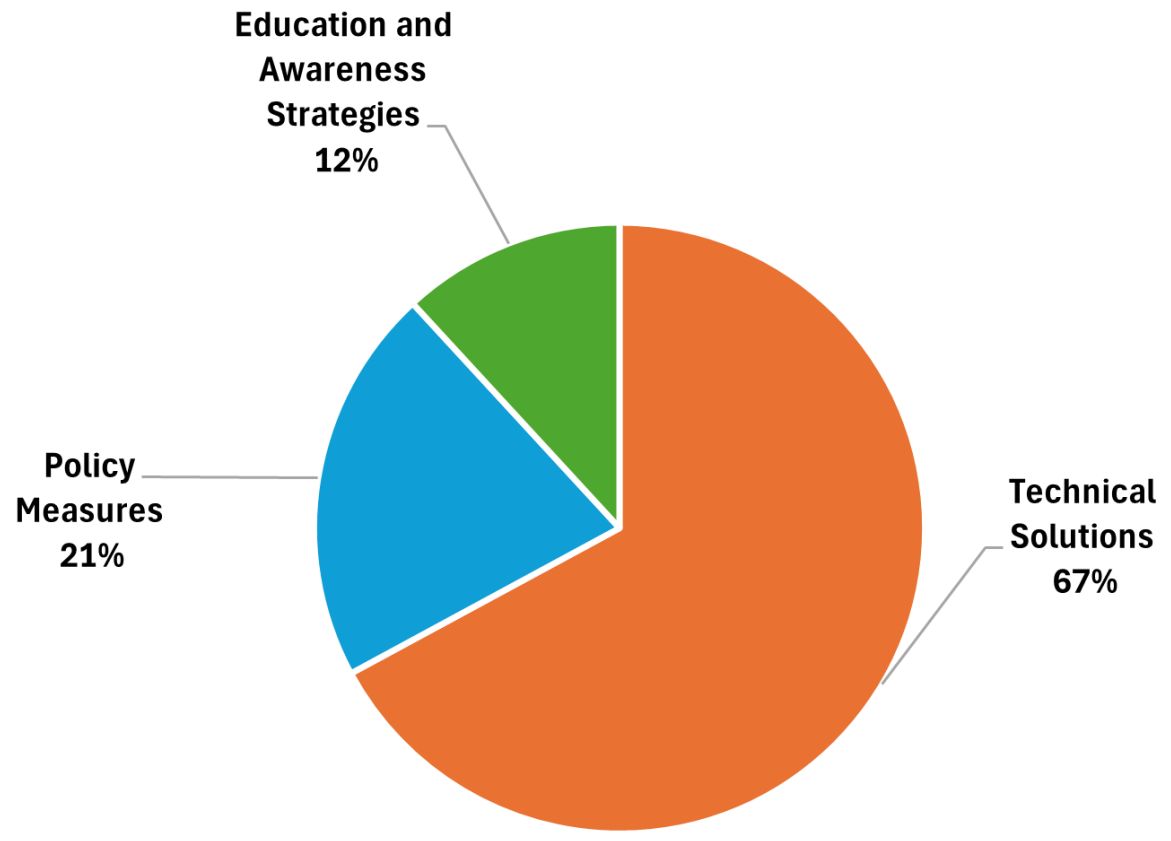

Fig. 2. Distribution of Papers Across Thematic Categories

For lightweight protection on constrained devices, surveys compare AES variants and lightweight cryptography (LWC: PRESENT and GIFT) [54], [55], [56], [57], while hybrid encryption prioritizes speed and NIST compliance. Cloud-IoT models and four-factor authentication (HE, biometrics, FL) advance efficiency and assurance [58], [59]. Specialized schemes include randomized non-overlapping image encryption via chaotic systems for cameras/wearables [60], ultralightweight RRSC-v1 sensor authentication [61], and hyperchaotic S-boxes with blockchain for tamper-proof media [62]. Attribute-Based Encryption (ABE) with fully hidden policies and fog offloading enables scalable access control [63], [64]; traceable ABE with blockchain supports auditable sharing [65], [66], [67]. Blockchain frameworks detect malicious nodes (up to 99.5% accuracy) [68], [69] and support searchable encryption (BC-RFMS; fuzzy multi-keyword) [70], [71], [72]. Trusted Execution Environments (TEEs: Intel SGX, ARM TrustZone) add hardware-level isolation for home data [73].

Human-centric and edge designs reduce exposure while maintaining usability. Emotion-aware smart home models [74] and ToyUI tools [75] address youth-specific needs, while on-device processing techniques like PriMask [76] and edge-assisted homomorphic encryption reduce privacy risks [77], [78], [79]. The CNN-BiLSTM intrusion detection system in [80] exemplifies edge-optimized security, detecting attacks in real-time with minimal latency, a crucial feature for youth devices like smartwatches or educational IoT platforms. Location privacy is enhanced via dummy-based anonymization (Enhanced-DLP, 21) and ring-loop routing [81], and adaptive methods like SDASL counter fingerprint-based snooping attacks in smart homes [82], [83], [84], [85]. Emerging technologies show promise, including zero-trust architectures with continuous authentication [86], cognitive digital twins with privacy-preserving FL [87], [88], and AI-driven security frameworks [26], [89]. However, challenges persist in scalability for resource-constrained devices [17], [37], [90], policy-implementation gaps [18], [91], and the need for enhanced digital literacy programs [26], [92]. Collectively, these advances form an evolving toolkit for youth privacy across wearables and educational IoT, while highlighting priorities for future research and deployment [93], [94].

### B. *How do policy frameworks address youth privacy risks?*

Policy effectiveness varies across jurisdictions. In Canada, PIPEDA has been criticized for its lack of explicit protections for youth, relying instead on generic consent models that fail to account for minors' unique vulnerabilities, as highlighted in studies examining its application to facial recognition and smart devices [5], [10], [28]. This aligns with findings from [31], where youth aged 13-16 expressed frustration with incomprehensible privacy policies and advocated for interactive consent tools like toggle switches and just-in-time notices and [3]'s critique of "educational exception" in IoT deployments, where schools bypass consent under the guise of pedagogy, creating loopholes in Children's Online Privacy Protection Act (COPPA) and GDPR compliance. While proposed reforms like the Consumer Privacy Protection Act (CPPA; Bill C-27) aim to strengthen consent requirements, they still fall short of providing robust, youth-specific safeguards [5], [32]. In contrast, the EU's GDPR and the U.S.'s COPPA impose stricter controls (e.g., parental consent under 16; prohibitions on profiling under 13) [29], [95], [96], yet enforcement lags and youth over 13 fall into regulatory gray zones [33], [1], [97]. The UK's Age-Appropriate Design Code (privacy-by-default) and the U.S. Kids Online Safety Act (algorithmic transparency) are steps forward, but unevenly adopted and overly reliant on parental controls [5], [22], [98].

Technical shortcomings compound policy gaps: youth-oriented devices often exhibit weak encryption, leakage, and covert surveillance [16], [99]. Blockchain and homomorphic encryption show potential, e.g., privacy-preserving aggregation for smart grids with lower computation and ~80% detection accuracy [100] and blockchain-enabled IoT forensics (PBCIS-IoTF) for integrity and traceability [101]. Vulnerabilities persist in educational VR [93], and functionality commonly trumps security [16], [102]. Despite the clear value of data minimization and on-device processing, PbD remains underused; proposals like SANIJO embed socio-ethical policies but see limited uptake [21], [96], [102].

Education and collaboration are critical to bridging these gaps. Digital literacy should move beyond "notice and consent" [1], [32], while interoperability standards enable safer device ecosystems [103]. Participatory approaches involving youth improve relevance and uptake [1], [97]. Progress requires stricter mandates (e.g., encryption, privacy-by-design), manufacturer accountability, and curricular integration of privacy literacy [16], [91], [32], [98]. A harmonized, multi-layered approach (legal, technical, and educational) with cross-border coordination is consistently recommended [95], [98]. Future policies must also grapple with emerging technologies like AI-driven surveillance and biometric data collection, which pose new threats to youth privacy [10], [32]. As [104] warns in their IoT security review, without urgent action, youth remain vulnerable to evolving data exploitation tactics. By addressing these challenges holistically, stakeholders can create a safer digital environment for young users, balancing innovation with ethical responsibility [96], [97].

### C. *What role do educational strategies play in enhancing youth awareness and control over their data?*

Education is pivotal, focusing on curriculum integration, multi-stakeholder collaboration, and empowerment. Formal education initiatives show promise, as [105] documented successful EU programs using gamified tools like privacy comics, while [19] found hands-on technology courses improved both technical skills and privacy awareness. These findings align with work on child-computer interaction tools that teach privacy-by-design principles through play [75].

Nonetheless, baseline gaps are large (e.g., 85% of young gamers are unaware of IoT risks) and scale-up across cultures is difficult [16], [48]. Fog-based architectures can support deployments [106], while crowdsensing projects teach privacy-preserving methods [77]. Regulatory frameworks should mandate youth-inclusive privacy education [32], [33], and industry–education partnerships aid curriculum co-creation [19], [107], aligned with practical IoT use in schools [30]. Empowerment-based and participatory models engage youth directly [1], [108]; research on privacy fatigue underscores the "privacy paradox," where concern does not translate into behavior [51], [109]. Real-time nudges show promise in closing this gap [6], [108], [110]. Persistent challenges (technical/cultural barriers) [16], [48], policy shortcomings [1], [32], and weak evaluation [6], can be addressed through cross-disciplinary privacy education modules [19], [105], culturally attuned tools [74], [75], robust impact measurement frameworks [16] and supportive architecture [106], [111].

### *D. What challenges exist in implementing these privacy measures at scale?*

Implementing privacy measures at scale for youth-focused smart devices faces multiple technical, policy, and educational challenges. Resource-constrained IoT devices hinder consistent deployment of robust protections [106], [112], [113]. While lightweight options like dynamic AES-CBC and FL help [20], [114], but introduce trade-offs (e.g., +50 ms latency, -10% throughput in real networks) [115]. Centralized storage creates breach risks, motivating decentralized and fog/blockchain architectures [42], [114], [116], though surveys note scalability and interoperability gaps [117]. Heterogeneous ecosystems complicate standardization [68], and many proposals lack empirical validation [109], [118], [119].

From a policy perspective, challenges include inadequate enforcement mechanisms within existing frameworks like PIPEDA, often leaving youth-specific privacy concerns insufficiently addressed and resulting in inconsistent compliance [5], [13]. Consent-centric models are ill-suited to youth, and even promising age-appropriate controls remain partial [5], [10], [30]. Advanced analyses (e.g., opacity-based state estimation) highlight modeling needs beyond current regulations [120].

From an educational perspective, limited literacy initiatives and sparse curricular integration impede informed decision-making [16], [105], [107], though studies suggest that gamified tools could help [105]. These educational limitations reduce the capacity of youth to exercise informed privacy decisions across smart devices and are compounded by the "privacy paradox," where awareness interventions show short-term effects but fail to sustain behavioral change [110]. The sophistication of state-of-the-art defenses, such as CNN-BiLSTM intrusion detection system [80], widens the gap between available protections and user understanding, amplifying the need for sustained, scalable education.

### *E. How can policymakers, manufacturers, and educators work together to strengthen youth privacy protections?*

Effective protection of youth privacy requires a coordinated multi-stakeholder approach. Policymakers must address PIPEDA's gaps for teens [5], [13], by adopting GDPR-like age-appropriate design principles [1], [95], and stricter rules for IoT surveillance [10], informed by the EU AI Act [97], and Canada's Bill C-27 [5], and by cross-border data governance guidance [3], [121]. Privacy-by-design mandates should be strengthened, with youth participation via advisory boards [30], [1], [108].

Manufacturers must prioritize lightweight encryption [34], [114], [122], decentralized architectures like FL [26], [42], [123], blockchain-based access control [65], and zero-trust models [86], to secure IoT devices [16], [104]. Standardized, empirically validated security patterns remain a critical need [124] in IoT systems, particularly to address gaps in privacy-specific designs and scalability. User-centric designs, such as privacy meta-assistants for smart homes [125], dummy-based location privacy [126], and ABE [34], [65], can empower youth, alongside ethical AI and differential privacy in smart toys [75], [76].

Educators should implement engaging digital literacy programs that teach practical privacy skills, using methods like gamified learning and interactive workshops [105]. Curricula should address critical topics including AI ethics, smart device risks, and algorithmic awareness [30], [95], [127]. The approach must include both youth education and parental guidance while being adaptable to different cultural contexts [107].

Policymakers and manufacturers should collaborate on security standards like ISO 27001 [127], while educators and technologists co-develop practical solutions such as privacy-preserving wearables [107]. These efforts work best when global standards are adapted locally, like implementing digital literacy programs in culturally relevant ways [105].

Table 1 summarizes the key findings and distribution of references across the five research questions. The discussion section synthesizes key insights, explores implications for policy, education, and industry, and identifies limitations in current approaches to protecting youth privacy in AI-driven smart devices.

## V. DISCUSSION

### *A. Privacy-by-Design in Technical Systems*

Our findings show that while numerous technical privacy-by-design mechanisms, such as FL, on-device processing, blockchain integration, and lightweight encryption, demonstrate substantial promise in minimizing data exposure, their deployment across youth-focused smart devices remains inconsistent. For instance, FL frameworks like the PoAh-enabled architectures achieve 99% accuracy in intrusion detection [39], and hybrid homomorphic encryption models offer efficient cloud-IoT security [58]. Yet, barriers persist, including device-level resource limitation, system interoperability issues, and the lack of standardized design protocols across manufacturers. These challenges inhibit large-scale adoption, despite widespread consensus in the literature that local processing and decentralized architecture offer strong protection for sensitive youth data. Furthermore, while model fusion techniques and secure inference frameworks offer granular data control, they remain inaccessible to mainstream device ecosystems without significant infrastructure investment or industry alignment, as seen in the slow adoption of NIST-compliant lightweight cryptography [54].

TABLE I. SUMMARY OF KEY THEMES AND RESEARCH FINDINGS

| Key Findings | References |
|---|---|
| **RQ1.** Technical solutions like Federated Learning (FL), blockchain-enhanced FL (BCFL), lightweight encryption (AES variants, LWC), and Trusted Execution Environments (TEEs) enhance youth data security. Privacy-preserving methods such as differential privacy, certificateless authentication, and hybrid encryption are critical for youth-focused applications. Edge computing and human-centric designs further improve security. | [16], [32]-[48], [49]-[57], [58]-[70], [71]-[82], [83]-[91] |
| **RQ2.** Policy frameworks vary in effectiveness, with PIPEDA lacking explicit youth protections, while GDPR and COPPA offer stricter safeguards. Challenges include enforcement gaps, inconsistent global adoption, and over-reliance on parental controls. Emerging solutions include blockchain-based forensics, privacy-by-design principles, and participatory policy frameworks. Educational and technical strategies are needed for comprehensive youth privacy protection. | [1], [3], [7], [8], [12], [14], [17], [18], [22]-[28], [88], [90], [92]-[102] |
| **RQ3.** Educational strategies (curriculum integration, gamification, stakeholder collaboration) improve youth privacy awareness.<br>Empowerment-based models (participatory design, real-time nudges) help bridge the privacy paradox. Challenges include knowledge gaps, scalability across cultures, and policy shortcomings.<br>Multi-stakeholder approaches (industry-education partnerships, fog-based learning) enhance effectiveness. | [5], [8], [12], [15], [25], [27], [43], [46], [71], [72], [103]-[109] |
| **RQ4.** Implementation challenges include technical (resource constraints, compatibility issues), policy (weak enforcement, vague consent models), and educational (low digital literacy, privacy paradox) barriers. Lightweight encryption and decentralized architectures help, but scalability and interoperability remain issues. Structured educational programs and empirical validation of solutions are needed. | [3], [7], [10], [12], [13], [16], [25], [37], [65], [104], [105], [107], [108], [110]-[118] |
| **RQ5.** Policymakers should adopt GDPR-like age-appropriate design principles and strengthen IoT regulations. Manufacturers must prioritize lightweight encryption, federated learning, and ethical AI. Educators should implement engaging digital literacy programs. Multi-stakeholder collaboration is essential for scalable, culturally adaptable solutions. | [1], [3], [7], [8], [10], [12], [22], [25], [29], [37], [62], [83], [92], [93], [95], [102], [103], [105], [119]-[125] |

## B. *Education and Digital Literacy as a Missing Pillar*

Despite growing recognition of youth empowerment, education is underused and unevenly implemented. This imbalance is reflected in the corpus, where only 12% of studies focus on education and awareness (Fig. 2). Studies reveal that 85% of young gamers lack awareness of IoT risks [16], and teens struggle with incomprehensible privacy policies [31]. Tools like gamified learning modules, participatory design approaches, and school-industry partnerships have shown clear benefits in raising youth awareness, yet their adoption in formal education systems is still limited, often due to resource gaps or policy mandates [75]. Moreover, the "privacy paradox", where young people express concern about privacy but still engage in risky data-sharing behaviors remains a persistent challenge, potentially linked to psychological barriers like privacy fatigue [109]. Educational strategies that use behavioral nudges, real-time feedback, and hands-on learning show promise in closing this gap between awareness and behavior. However, scaling these efforts requires localized program design and consistent evaluation frameworks to measure long-term impact [16].

## C. *Toward a Multi-stakeholder Framework*

The review reinforces the importance of a coordinated multi-stakeholder approach involving policymakers, manufacturers, and educators. Policymakers must go beyond passive frameworks like PIPEDA, which lacks explicit youth protections, and proactively legislate youth-specific mandates, such as the UK's Age-Appropriate Design Code's privacy-by-default rule [5]. Manufacturers should embed privacy-by-design principles across the device lifecycle, adopting lightweight encryption and decentralized architecture, while collaborating with regulators to certify compliance. Finally, educators must lead in developing cross-disciplinary privacy literacy programs, such as teaching AI ethics through privacy-preserving wearable projects [107], and co-creating culturally attuned tools with youth advisory boards [1]. Without such alignment, youth privacy will remain compromised in an increasingly connected device ecosystem. Given that policy-focused work constitutes 21% of the corpus, coordinated policy-industry-education initiatives are needed to correct this skew (Fig. 2).

## D. *Recommendations*

To address the gaps and challenges identified in the review, we propose actionable recommendations for policymakers, manufacturers, and educators. First, technical implementation must prioritize standardized privacy-by-design and youth-centric solutions, such as lightweight encryption and a federated learning framework, to accommodate resource-constrained devices. Decentralized models like blockchain-enhanced architecture should be adopted to ensure auditable data integrity in smart ecosystems, particularly for high-risk applications like education and healthcare IoT. Second, policymakers, reforming PIPEDA is critical to address its current gaps in youth-specific safeguards. While PIPEDA currently relies on generic consent models, amendments should mandate age-appropriate design like those in the EU's GDPR and the UK's Age-Appropriate Design Code by requiring default privacy settings, minimal data collection and transparent algorithmic governance for youth-adopted devices. Proposed legislation, such as Bill C-27 (CPPA), should explicitly integrate privacy-by-design obligations for manufacturers, ensuring encryption, decentralized architecture, and third-party audits for compliance. Cross-sectional collaboration with provinces and international partners could help to standardize regulations, particularly for devices with cross-border data flows. Third, PIPEDA should be revised to include mandatory privacy literacy programs in schools, incorporating gamified tools and hands-on workshops on smart-device risks. Partnerships with manufacturers could pilot privacy-by-design driven curricula, such as teaching federated learning concepts or secure authentication practices. By aligning PIPEDA reforms with technical privacy-by-design standards and proactive education, Canada can model a youth-centric privacy framework that balances innovation with protection.

### E. Limitations

This review is restricted to English-language articles, which may limit the generalizability and exclude non-Western innovations. Time constraints on the project and the volume of available information may also have led to incomplete reviews. Additionally, while this study identifies promising technical solutions, many lack large-scale empirical validation in youth-specific settings, highlighting a critical gap for future research. Furthermore, the rapid evolution of AI-IoT devices means that literature may not fully cover the latest innovations. This study should be viewed as a snapshot in time rather than a definitive overview of the literature.

## VI. Conclusion

This systematic review provides a comprehensive evaluation of privacy-by-design frameworks, tools, and educational interventions protecting youth privacy in AI-enabled smart devices. Drawing from 2,216 sources and culminating in the analysis of 122 studies, the review reveals that while significant progress has been made in developing technical and policy-based privacy solutions, substantial gaps remain in implementation, enforcement, and youth engagement. The findings demonstrate that federated learning, on-device inference, and lightweight cryptographic protocols can effectively minimize data exposure, yet their scalability is hindered by hardware limitations and a lack of standardization. Policy frameworks such as PIPEDA lack youth-specific enforcement mechanisms, contrasting sharply with more prescriptive models like the EU's GDPR and the UK's Age-Appropriate Design Code. Educational initiatives, while impactful in isolated studies, are rarely integrated into formal curricula or evaluated across diverse sociocultural contexts. This study highlights the urgent need for a multi-stakeholder approach, uniting policymakers, educators, and manufacturers to co-develop scalable, context-sensitive, and youth-centered privacy ecosystems. Youth deserve technologies that do not force trade-offs between functionality and safety, but instead embed privacy, ethics, and empowerment into their core design. Moving forward, researchers must address critical gaps through longitudinal and cross-cultural studies, evaluate the real-world efficacy of privacy-enhancing technologies, and develop adaptive frameworks that evolve alongside smart device innovation. Ultimately, the goal must be to ensure that smart devices are not only advanced but also safe, equitable, and respectful of young people's fundamental right to privacy.

## Acknowledgment

This project has been funded by the Office of the Privacy Commissioner of Canada (OPC); the views expressed herein are those of the authors and do not necessarily reflect those of the OPC.